\documentclass[%
 reprint,
 amsmath,amssymb,
 aps,
]{revtex4-2}

\usepackage{graphicx}
\usepackage{dcolumn}
\usepackage{bm}
\usepackage{bbm}
\usepackage{bbold}

\renewcommand{\d}[1]{\ensuremath{\operatorname{d}\!{#1}}}

\def\std{^{\ensuremath{-\kern-5pt{\ominus}\kern-5pt-}}}

\renewcommand{\v}{\boldsymbol}

\begin{document}

\preprint{APS/123-QED}

\title{Pumping current in a non-Markovian $N$-state model}

\author{Ville M. M. Paasonen}
 \altaffiliation[E-mail: ]{ville.paasonen@yukawa.kyoto-u.ac.jp}
\author{Hisao Hayakawa}%
\affiliation{%
 Yukawa Institute for Theoretical Physics, Kyoto University\\
 Kyoto, 606-8502, Japan}%
 
 \date{\today}

\begin{abstract}
A periodically modulated $N$-state model whose dynamics is governed by a time-convoluted generalized master equation is theoretically analyzed. It is shown that this non-Markovian master equation can be converted to a Markovian master equation having a larger transition matrix, which affords easier analysis. The behavior of this model is investigated by focusing on the cycle-averaged pumping current. In the adiabatic limit, the geometrical current is calculated analytically, and compared to numerical results which are available for a wide range of modulation frequencies. 

\end{abstract}

\maketitle


\section{\label{intro}Introduction}

Master equations (MEs) are widely used in non-equilibrium statistical mechanics to model the time evolution of a range of classical and quantum mechanical systems. The mathematical foundation of MEs is the differential Chapman-Kolmogorov equation of stochastic analysis \cite{gardiner2009stochastic}, and in its basic form, it only describes systems that carry no memory of their past: this property is referred to as Markovianity. Is is known, however, that due to a number of physical reasons, real systems do usually possess some degree of memory of their past evolution, and thus obey a non-Markovian ME (nMME). An archetypal example of this is the fluctuation-dissipation relation \cite{Kubo_1966}, which shows that any time-nonlocal correlations in the environment necessarily lead to memory effects. Theoretically, non-Markovianity has been attracting attention in classical mechanics as a link between continuous-time random walks and time convoluted MEs \cite{ClGME, PhysRevE.99.062137}.  In quantum physics, the connection between the flow of information and non-Markovian processes, as well as quantum measurements of Markovianity have received considerable research effort \cite{PhysRevA.86.012115, RevModPhys.88.021002, PhysRevLett.103.210401}. Furthermore, advancements in experimental techniques have made it possible to directly measure non-Markovianity in the context of classical \cite{bexpt} and quantum \cite{qopt, ion} systems.

Modulating control parameters such as rate constants, bath temperatures or gate voltages of a physical system out of equilibrium can lead to net flow of a physical current, e.g. flow of product, heat flow or electron current, even in the absence of a net bias of the control parameters \cite{PhysRevB.27.6083}. The origin of the  pumping current is the Berry-Sinitsyn-Nemenman (BSN) phase, which was originally discovered in the context of quantum systems \cite{10.2307/2397741, Sinitsyn_2007, PhysRevLett.99.220408}. Geometrical current is generated when the modulation speed of the parameters is sufficiently slow, i.e., in the adiabatic limit. The BSN phase has a significant impact on quantum transport \cite{PhysRevLett.104.170601, PhysRevB.86.235308}, and it also leads to a path-dependent geometrical entropy \cite{PhysRevE.84.051110, JStatPhys}. The effects of the BSN phase have been extensively studied in the spin-boson model \cite{10.1093/ptep/ptu149}. Over the recent years, effects of finite speed modulation, i.e., non-adiabatic effects, have also been investigated for this model \cite{PhysRevE.89.052108}. Furthermore, it has been shown that the presence of the BSN phase engenders non-Gaussianity of the system fluctuations, leading to a modified form of the fluctuation theorem for geometrical pumping \cite{PhysRevE.96.022118, hino2019fluctuation}. Moreover, a heat engine using Thouless pumping \cite{PhysRevB.27.6083} associated with the BSN curvature has been theoretically designed \cite{PhysRevResearch.3.013187}.

The essential features of this geometrical effect  have also been shown to exist in classical systems, such as the Sinitsyn-Nemenman (SN) model of reaction kinetics, in which some parameters are cyclically modulated by an external agent \cite{Sinitsyn_2007, PhysRevLett.99.220408}.  Recently, the adiabatic result has been extended to the non-adiabatic regime also for this model \cite{fujii2019nonadiabatic}. It was found that the pumping current reaches a peak after the adiabatic regime in which the current linearly increases with the modulation frequency, and eventually decays as the inverse of the modulation frequency in the asymptotic limit. Finite modulation speed means that the pumping current can no longer be expressed using strictly geometrical quantities, but a formally geometrical expression in terms of a line integral in parameter space is still possible. Furthermore, the effect of non-adiabaticity on the fluctuation theorem has been investigated in the context of the SN model \cite{newPub}.  Motivated by the attention these recent results have garnered, in the present paper a non-Markovian generalization of the SN model is presented as a natural extension of the aforementioned research, and its  adiabatic and non-adiabatic behavior is investigated analytically and numerically.

The structure of this paper is as follows: In Sec. \ref{nMSN}, we first introduce our Markovian $N+4$ model and then perform a tracing out procedure to obtain an $N$-state non-Markovian system. The results obtained for the pumping current are shown in Sec. \ref{results}. We obtain the geometrical form of the adiabatic current in the low frequency limit and compare this to the full non-adiabatic current obtained using numerical methods. In Sec. \ref{discussion}, the results obtained and their physical implications are discussed, and a conclusion is presented in Sec. \ref{conclusion}.  Supplementary information and details of the various calculations in the main text will be presented in the Appendices.

\section{\label{nMSN}Non-Markovian Sinitsyn-Nemenman Model}

In this section, we introduce our non-Markovian model by deriving it from a Markovian model having a larger transition matrix. This procedure is analogous to the Nakajima-Zwanzig method in which we obtain a system obeying non-Markovian dynamics by tracing out degrees of freedom \cite{Chaturvedi1979, Shibata1977, doi:10.1143/JPSJ.49.891}. However, as will be discussed below, we can solve the Markovian model with a larger matrix in some situations, because the larger Markovian system is phenomenological and governed by  simple dynamics. 
Therefore, in contrast to the relation between the full von-Neumann equation and the Redfield equation or Lindblad equation for a subsystem in the Nakajima-Zwanzig method, in our case the relation between the Markovian model and non-Markovian model is simple, affording exact conversion between them.

\subsection{Markovian Formulation}

Let us consider a Markovian master equation for a system with $N + 4$ states,
\begin{equation}
\frac{\d{}}{\d{t}} \rho _i(t) = k_0 \sum _{j=1}^{N+4}L_{ij}(t) \rho _j(t)\text{,}\label{1st}
\end{equation}
where $\rho _i$ is the probability of state $i$ and the element $L_{ij}$ of the transition matrix is the hopping rate between states $i$ and $j$.  For convenience, we have introduced the inverse time scale (characteristic hopping rate) $k_0$ so that all other quantities except for time $t$ become dimensionless. The time dependence of the hopping rates arises from modulation by an external agent. This can be realized for  instance by controlling the temperatures or chemical potentials of the environments, though the temperature control is not easy in realistic situations \footnote{Nevertheless, it is remarkable that the effective temperature of a Stirling engine can be controlled by a sinusoidal function (see e.g. Ref. \cite{PhysRevE.102.012142}).}. 
In the following, the rates will be assumed to undergo cyclic modulation in time.
We also assume that the environments attached to the system are large enough such that the back-reaction from the system is negligible and the probability distribution in each environment is always described by an equilibrium distribution $\rho _i ^{\nu, \rm eqm}$, where $\nu=\mathrm{L}$ or $\mathrm{R}$ is the index to specify the left (L) or right (R)  environment attached to the system,  even when some parameters such as the temperature and chemical potential are controlled by an external agent. 

We first briefly review the properties of the ME required for conservation and non-negativity of the probabilities as well as detailed balance to hold.
For probability to be conserved, the matrix elements must satisfy \cite{gardiner2009stochastic}
\begin{equation}
\sum _{i=1}^{N+4} L_{ij} = 0, \quad \forall j \label{probcon} \text{.}
\end{equation}
The condition for complete positivity is more restrictive: we are dealing with a continuous process in time, which means that the only way for a probability, say $\rho _i$, to become negative is if it first passes through zero. We thus require \cite{gardiner2009stochastic}
\begin{equation}
\left. \frac{\d{}}{\d{t}} \right|_{ \rho _i =0} \rho _i = \sum _{j\neq i} L_{ij} \rho_j \geq 0
\end{equation}
for arbitrary non-negative $\rho _i$. This in turn means that all off-diagonal elements of $L$ ought to be non-negative. 

Finally, we impose detailed balance. The principle of detailed balance states that a process and its reverse happen at the same rate at equilibrium:
\begin{equation}
L_{ij}\rho^\mathrm{eqm}_j = L_{ji}\rho^\mathrm{eqm}_i\text{.}
\end{equation}
The equilibrium probability distribution would usually have the Boltzmann form characterized by the inverse temperature $\beta$ and energy $E_i$ of the state $i$, i.e., $\rho^\mathrm{eqm}_i \propto e^{-\beta E_i}$ so that the above equation can be written as
\begin{equation}
L_{ij}e^{-\beta E_j} = L_{ji}e^{-\beta E_i}\text{.}
\end{equation}
Let us explore what this means for our ME. Summing over $j$, we obtain
\begin{equation}
\sum _j L_{ij}\rho^\mathrm{eqm}_j = \sum _j L_{ji}\rho^\mathrm{eqm}_i = 0 \text{,}
\end{equation}
where we used Eq. \eqref{probcon}. Thus, we arrive at 
\begin{equation}
\frac{\d{}}{\d{t}} \rho _i^\mathrm{eqm}(t) = \sum _{j=1}^{N+4}L_{ij}(t) \rho ^\mathrm{eqm} _j(t) =  0 \text{.}
\end{equation}
This is not only a statement of the intuitive notion that an equilibrium state should exist, but also an important condition to determine the hopping rule between the system and the environment. It can be guaranteed to hold by requiring
\begin{equation}
\det{L} = 0\text{.}
\end{equation}
Note that this condition is already guaranteed by Eq. \eqref{probcon}. Thus, requiring detailed balance to hold does not impose additional constraints.
We note that the Markovian SN model treated in e.g. Ref. \cite{Sinitsyn_2007} corresponds to a Markovian two-state model.
Now let us rewrite the Markovian ME in matrix notation,
\begin{equation}
\Omega  \frac{\d{}}{\d{\theta}} \v{\rho}(\theta) = L(\theta) \v{\rho}(\theta) \text{,}\label{timeT}
\end{equation}
where we have introduced the phase variable $\theta:=\Omega k_0(t-t_0)$ with the dimensionless angular frequency $\Omega$ of cyclic modulation and the time $t_0$ required to reach a cyclic state which is independent of a specific choice of initial conditions. We assume that the system can be partitioned into one $N$-state site and an adjacent two-state site on each side:
\begin{equation}
\v{\rho} (\theta)=
\begin{bmatrix}
\v{q}^\mathrm{L}(\theta)\\[0.5em] 
\v{p}(\theta)\\[0.5em] 
\v{q}^\mathrm{R}(\theta) \\[0.5em] 
\end{bmatrix} \text{,}
\end{equation}
i.e. the $N$-state reduced system $\v{p}$ and the left and right adjacent states $\v{q}^\mathrm{L}$ and $\v{q}^\mathrm{R}$, respectively. We note here that this partitioning can be considered analogous to the projection operators $\mathcal{P}$ and $\mathcal{Q}$ of the Nakajima-Zwanzig approach, though the sizes of adjacent states $\v{q}^\mathrm{L}$ and $\v{q}^\mathrm{R}$ are small in our model. 

Provided that the adjacent states do not couple to each other directly, the ME becomes
\begin{equation}
\Omega  \frac{\d{}}{\d{\theta}} \v{\rho}(\theta) =  
\begin{bmatrix}
K^\mathrm{L} (\theta)& W_\mathrm{out}^\mathrm{L} (\theta)&0 \\[0.5em] 
  W_\mathrm{in}^\mathrm{L}(\theta) & S(\theta) &  W_\mathrm{in}^\mathrm{R}(\theta)\\[0.5em] 
0& W_\mathrm{out}^\mathrm{R} (\theta) &K^\mathrm{R}(\theta)  \\[0.5em] 
\end{bmatrix}
\begin{bmatrix}
\v{q}^\mathrm{L}\\[0.5em] 
\v{p}\\[0.5em] 
\v{q}^\mathrm{R} \\[0.5em] 
\end{bmatrix}\text{.}\label{subMatForm}
\end{equation}
Here, $S(\theta)$ is a $N\times N$ square matrix, $ W_\mathrm{in}^\nu(\theta) $ are $N\times 2$ and $ W_\mathrm{out}^\nu(t) $ are $2\times N$ rectangular matrices, and $K^\nu (\theta)$ with $\nu = \mathrm{L}, \mathrm{R}$ are $2\times 2$ square matrices. This small matrix $K^\nu(\theta)$ enables us to understand the relationship between non-Markovian dynamics in a subsystem and Markovian dynamics in the full system. 

Next, we specify the structure of the submatrices introduced above.  Firstly, $K^\nu(\theta)$ are assumed to be given by
\begin{align}
K^\nu (\theta) &= 
\begin{bmatrix}
- k_\mathrm{2} ^\nu (\theta) - w &k^\nu_\mathrm{1}(\theta) \\[0.5em] 
k^\nu_\mathrm{2}(\theta) &-k_\mathrm{1} ^\nu (\theta)- w\label{Ksplit}
\end{bmatrix},
\end{align}
where, to make the model concrete, we choose \cite{fujii2019nonadiabatic}
 \begin{equation}
\begin{aligned}
 k_{1}^\mathrm{R}(\theta) &= k_{2}^\mathrm{L}(\theta)= k_1(\theta)  \text{,}\\
k_{1}^\mathrm{L}(\theta) &= k_{2}^\mathrm{R}(\theta) =k_2(\theta)   \text{,}\\
k_1 (\theta) &=   1 + r\cos \theta    \text{,}\\
k_2(\theta) &=   1 + r\cos ( \theta -\phi )   \text{,} \label{param}
\end{aligned}
\end{equation}
where the internal rate constant $w>0$, amplitude $r$, phase difference $\phi$ and dimensionless modulation frequency $\Omega$ of Eq. \eqref{timeT} are used as control parameters. This choice can be motivated as follows. Firstly, in the space of the rate coefficients, the protocol trajectory must enclose a finite area for finite pumping current to be observed in the adiabatic limit \cite{fujii2019nonadiabatic}. The phase difference $\phi$ in the last equation of Eq. (13) ranges from 0 to $\pi/2$. Below we mainly focus on the choice $\phi = \pi/2$ which gives  a circle of radius $r$ centered at $k_1 =1, k_2=1$, but for  illustrative purposes we will also consider a straight line which does not enclose any area, obtained by setting $\phi$ to zero. Secondly, as can be seen from Appendix \ref{6x6Details}, to obtain a finite pumping current, there needs to be a phase difference between the left and right rates as well as the rates of the two states on each side. 

We note that the concrete calculations below are performed for $N=2$, where $S$, $W_\mathrm{in}^\nu$ and $W_\mathrm{out}^\nu$ were assumed to be proportional to the $2\times 2$ identity matrix $\mathbb{1} $: $S=-2w\mathbb{1}$, $W_\mathrm{in}^\nu = W_\mathrm{out}^\nu = w \mathbb{1}$, though we do not specify $N$ in the general framework. The full transition matrix for $N=2$ is shown explicitly in Eq. \eqref{fullglory}.

As can be checked by direct calculation, the Markovian equation \eqref{subMatForm} together with the above structure of the submatrices guarantees the conservation and non-negativity of $\v{\rho}$, thus also satisfying detailed balance. Physically, an important feature of this model is that the two states at each site, e.g. $q_1^\mathrm{R}$ and $q_2^\mathrm{R}$ for the right adjacent site, can only mix in the left and right environments, so there is no coupling between them within the system. Furthermore, we assume that only $ k_{1}^\nu$ and $ k_{2}^\nu $ have time dependence. Physically, this corresponds to externally modulating only the coupling between the adjacent states and the environment.

We note that $K^\nu$ in Eq. \eqref{Ksplit} can be split as
\begin{equation}
K^\nu (\theta) = K_0 +  r K_1 ^\nu (\theta) \text{,}
\end{equation}
where, $K_0$ and $K_1 ^\mathrm{L} (\theta)$ are, respectively,
\begin{align}
K_0  = 
\begin{bmatrix}
-1 -w   &  1 \\[0.5em] 
1 &-1 - w  
\end{bmatrix}, \,
K_1 ^\mathrm{L} (\theta) =
\begin{bmatrix}
 -\cos \theta&  \cos  (\theta-\phi) \\[0.5em] 
\cos \theta & -\cos (\theta-\phi) 
\end{bmatrix}\text{,}
\end{align}
and $K_1 ^\mathrm{R} (\theta)$ is obtained by exchanging $\cos\theta$ with $ \cos  (\theta-\phi)$.
Furthermore, with the quantity of interest, the pumping current, in mind, we introduce the counting field $\chi$:
\begin{equation}
K^\mathrm{R} (\theta) \to K^\mathrm{R} (\theta, \chi)  := 
\begin{bmatrix}
-k_\mathrm{2} ^\mathrm{R} (\theta) -w  &e^{\chi} k ^\mathrm{R} _\mathrm{1}(\theta) \\[0.5em] 
e^{-\chi} k^\mathrm{R} _\mathrm{2}(\theta) &- k _\mathrm{1} ^\mathrm{R}  (\theta)-w 
\end{bmatrix}\text{.}
\end{equation}
We may now consider the modified ME,
\begin{equation}
\Omega  \frac{\partial}{\partial{\theta}} \v{\rho}(\theta, \chi) = L (\theta, \chi) \v{\rho}(\theta, \chi) \text{,}\label{modME}
\end{equation}
where the partial derivative is used to emphasize the dependence of $\v{\rho}$ on $\chi$. The modified transition matrix $L (\theta, \chi)$ is defined as the original matrix with $K^\mathrm{R} (\theta)$ replaced by $K^\mathrm{R} (\theta, \chi) $; its explicit form is shown in Eq. \eqref{fullglory}. Thus, the modified probability vector $\v{\rho}(\theta, \chi)$ is simply defined as the solution of the modified ME, Eq. \eqref{modME}. Solving for $\v{\rho}(\theta, \chi) $, computing the cycle averaged characteristic function $z(\theta)$, 
\begin{equation}
\overline{z( \chi)} :=  \frac{1}{2\pi}\int _0 ^{2\pi} z(\theta, \chi) \text{,}
\end{equation}
where
\begin{equation}
z(\theta, \chi) := \sum _i \rho_i(\theta, \chi)\text{,}
\end{equation}
and defining the cumulant generating function,
\begin{equation}
g(\chi) = \ln  \overline{z( \chi)} 
\end{equation}
will now give us access to all moments of the pumping current. It can be readily shown that the pumping current, defined as the first moment, 
 \begin{equation}
 \overline{J(\Omega)}  :=  \left. \frac{\partial }{\partial \chi} \right|_{\chi=0}g(\chi) 
 \text{,}
 \end{equation}
can equivalently be obtained by taking the cycle average of the instantaneous current:
\begin{equation}
\overline{J(\Omega)}  = \frac{1}{2\pi}\int_{0}^{ 2\pi}J (\theta) \d{\theta} \label{avgCurr} \text{,}
\end{equation}
where
\begin{equation}
J(\theta) =k_\mathrm{in1}^\mathrm{R}(\theta)q_2^\mathrm{R}(\theta) - k_\mathrm{in2}^\mathrm{R}(\theta)q_1^\mathrm{R}(\theta)\label{numCurrdef} \text{.}
\end{equation}
Physically, this corresponds to the current flowing from the right adjacent states to the right environment.
In the adiabatic limit, this pumping current has a geometrical formulation and can be obtained analytically, as shown in the next section.

\subsection{Non-Markovian Form}
It is well-known that memory effects arise when we trace out environmental degrees of freedom. Physically, when the environment is allowed to back react with the system, the environment will retain information about the past states of the system, which can affect its dynamics. 

We now demonstrate how this arises for our system. Specifically, we show that the $N + 4$  -state Markovian ME is equivalent to an $N$-state non-Markovian ME. First, we rewrite the Markovian ME, Eq. \eqref{subMatForm}, in the split form
\begin{equation}
  \begin{cases}
\displaystyle{ \Omega  \frac{\d{}}{\d{\theta}}\v{p}(\theta) }=  S \v{p}(\theta) + \sum _\nu W _\mathrm{in}^\nu \v{q}^\nu(\theta) \text{,} \\[0.5em]
\displaystyle{ \Omega   \frac{\d{}}{\d{\theta}}\v{q}  ^\nu (\theta)} =K^\nu (\theta) \v{q}^\nu (\theta) + W _\mathrm{out}^\nu \v{p} (\theta)  \text{.} 
  \end{cases}\label{best}
\end{equation}
Next, we formally solve for $\v{q}^\nu$,
\begin{equation}
\v{q}^\nu(\theta) = \frac{1}{\Omega } U^\nu(\theta) \int_0^\theta \d{\theta'}[U^\nu(\theta')]^{-1}W _\mathrm{out}^\nu \v{p}(\theta')\text{,}
\end{equation}
where transient terms are neglected, and the time evolution operator $U^\nu(\theta)$ obeys the differential equation
\begin{equation}
\Omega \frac{\d{}}{\d{\theta}} U^\nu(\theta) =  K ^\nu(\theta) U^\nu(\theta)\text{.}\label{Udiff}
\end{equation}
While $U^\nu(\theta)$ can be written with the help of the time ordered exponential operator,
\begin{equation}
U(\theta)  = \exp_\leftarrow \left[ -\frac{1}{\Omega } \int_{0}^\theta \d{\theta'}K^\nu(\theta')  \right]\text{,}
\label{UT}\end{equation}
this expression is not useful for concrete analysis. 
Nevertheless, if we restrict our interest to the case of small $r$, Eq. \eqref{UT} is reduced to an exponential function.  Accordingly, we expand $U^\nu (\theta)$ as
\begin{equation}
U^\nu (\theta) = \sum _{n=0}^\infty r^n U_n^\nu (\theta)\label{Ur}
\end{equation}
which, upon substituting into Eq. \eqref{Udiff} and solving at each order, leads to
\begin{align}
n=0: \quad U_0^\nu (\theta) &=  e^{\theta K_0/ \Omega}, \\
n\ge 1: \quad U_n^\nu (\theta) &=  \int_0^\theta\d{\theta'} e^{(\theta-\theta')K_0/ \Omega} K_{1}^\nu(\theta') U_{n-1}^\nu(\theta')\text{.}
\end{align}
Substituting this into the differential equation for $\v{p}$, we indeed obtain a non-Markovian ME of the form 
\begin{align}
\Omega \frac{\d{}}{\d{\theta}}\v{p} (\theta) =  S \v{p}(\theta) +\int _0^\theta  \d{\theta'}  M (\theta,\theta') \v{p}(\theta')\text{,}\label{nMForm}
\end{align}
where the memory kernel is given by
\begin{align}
\quad M(\theta,\theta') &:= \frac{1}{\Omega} \sum_\nu W _\mathrm{in} ^\nu (\theta) U^\nu(\theta)[U^\nu(\theta')]^{-1}W _\mathrm{out}^\nu(\theta') \text{.}
\end{align}
On the other hand, we may obtain coupled equations for   $\v{q}^\nu(t)$ instead, as
\begin{align}
\v{q}^\nu(\theta) &=  \int _0^\theta \d{\theta'}  U^\nu(\theta)[U^\nu(\theta')]^{-1} W_\mathrm{out}^\nu \nonumber \\ 
&\times \left\{ \int_0^{\theta'}\d{\theta''}e^{-(\theta'-\theta '')K_0/\Omega}\sum_\nu W_\mathrm{in} ^\nu \v{q}^\nu(\theta'')   \right\}\text{.}
\end{align}
We emphasize that we have not used any approximations in deriving these time-nonlocal equations, so the important properties of the original Markovian model, namely that $\v{\rho}$ remains non-negative and conserved throughout the time evolution, are retained. However, the effectiveness of the above procedure hinges on there being only a few adjacent states $\v{q}^\nu$ connecting the system to each environment. In the general case, as treated by the Nakajima-Zwanzig formalism, tracing out a large number of irrelevant degrees of freedom does not lead to an analytically workable equation.

In many situations, the memory kernel $M(\theta, \theta ')$ is described by a multiple exponential function.   In fact, it is known in the literature on Markovian embedding of non-Markovian processes that an exponential memory kernel  is the easiest case to treat analytically \cite{kanazawa1, kanazawa2}.  However, our model does not lead to a multiple exponential function, but a more complicated form due to the time dependence introduced by the external modulation and small adjacent vectors $\v{q}^{\mathrm{L}}$ and $\v{q}^{\mathrm{R}}$, and thus, treating this system anaytically is non-trivial.

The advantage of our approach is that there is no need to deal with the non-Markovian dynamics: we can simply solve the equivalent Markovian system because the eliminated adjacent systems are small. Working with the non-Markovian form of the equations does not present any mathematical or physical merit. Indeed, we will not employ the non-Markovian form of the dynamics in the treatment that follows. However, deriving the time-nonlocal form serves as an interesting example of how non-Markovianity arises naturally in the context of externally modulated systems.

There are a number of alternative ways to deal with nMMEs that have been explored in the literature. The most straightforward approach is to transform the nMME into Laplace space and solve the resulting algebraic equation there. However, it often happens that $M(\theta, \theta ')$ does not depend only on the phase difference (as is the case here), so that the convolution theorem cannot be utilized. This issue can be circumvented by performing a Taylor expansion of $M(\theta, \theta ')$ around one of the time variables so as to create terms which only depend on the difference between $\theta$ and $\theta '$ \cite{PhysRevLett.103.136801}. While this allows transformation into Laplace space, it is generally difficult to perform the inverse transformation explicitly. Furthermore, the resulting solution is in the form of an infinite series instead of the closed-form approach of the present paper.

A more general approach to deal with nMMEs is based on the Nakajima-Zwanzig projection operator technique \cite{breuer2002theory, 10.1143/PTP.20.948, doi:10.1063/1.1731409}. In essence, one first eliminates the environment dynamics (corresponding to $\v{q}^\nu$ of Eq. \eqref{best}) to obtain a nMME, and then performs an expansion in the system-environment coupling to achieve a time-covolutionless (TCL) equation of motion for the system \cite{Chaturvedi1979, Shibata1977, doi:10.1143/JPSJ.49.891}. Again, however, while a number of refined expansion protocols have been developed over the recent years \cite{PhysRevB.83.115416}, a closed-form time-local equation which is easy to work with numerically or analytically cannot be derived using this approach.

\section{\label{results}Results}

\subsection{Analytical expression of the adiabatic current}

\begin{figure}
\centering
\includegraphics[width=0.5\textwidth]{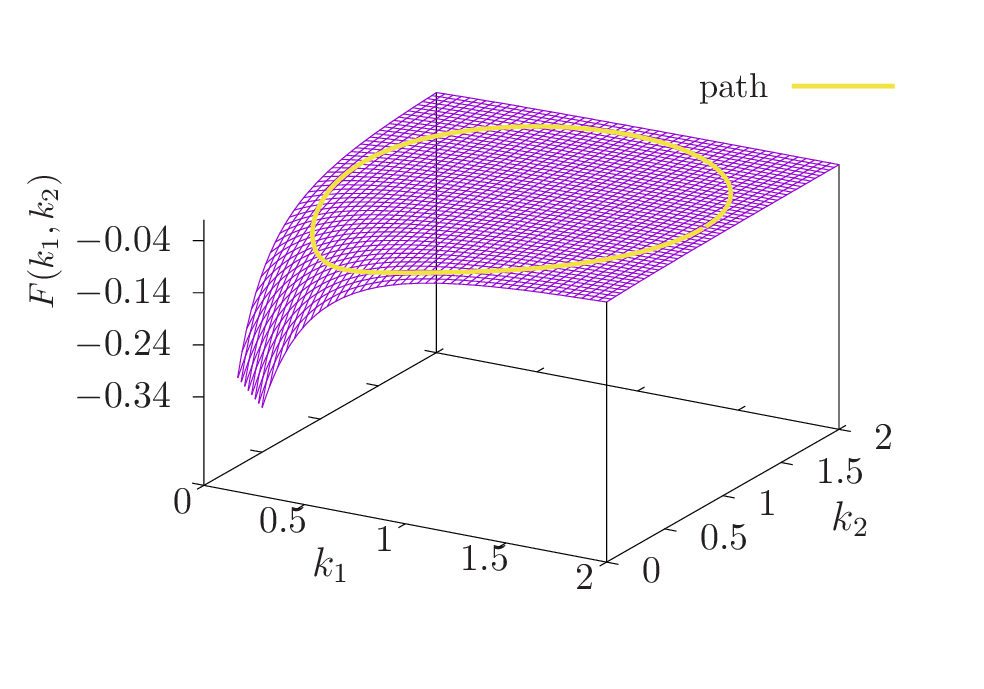}
\caption{\label{surf} The Berry curvature $F(k_1,k_2) $ for $w=2$ plotted together with the circular trajectory of the parameter modulation with $r=0.9$.}
\end{figure}

As can be seen from Eqs. \eqref{geom1} and \eqref{geom2} in Appendix \ref{6x6Details}, we can write the pumping current in terms of the Berry-Sinitsyn-Nemenman (BSN) curvature in the adiabatic limit:
\begin{align}
\overline{J_\mathrm{ad}(\Omega)}  = 
\iint \d{k}_1\d{k}_2 F(k_1(r,\theta), k_2(r,\theta)), 
\end{align}
where the BSN curvature is written as 
\begin{align}
F(k_1(r,\theta),k_2(r,\theta)) := \partial _{k_2} & A_1 (k_1(r,\theta),k_2(r,\theta)) \nonumber 
\\ &- \partial _{k_1} A_2 (k_1(r,\theta),k_2(r,\theta)) \nonumber \\
= -&\frac{w}{3[k_1(r,\theta)+k_2(r,\theta)+w]^3}\text{.} \label{BerryCurv}
\end{align}
The Berry vector potential $A_\alpha$ with $\alpha = 1,2$ is defined in terms of the eigenvectors of the modified transition matrix $L(\theta, \chi)$ in Eq. \eqref{geom2}. In the above equations, the rates $k_1$ and $k_2$ are treated as functions of the radial and azimuthal parameter which prescribe the area to be integrated over. Note that $F$ is a second-rank antisymmetric tensor in general, but in the case of only two independent modulation parameters, this can readily be reduced to a scalar quantity. A surface plot of the curvature together with the circular modulation trajectory (for $r=0.9$ and $\phi=\pi/2$) is shown in Fig. \ref{surf}. 

The parametrization of Eq. \eqref{param} has the Jacobian $|\sin \phi| r$, so we have
\begin{equation}
\overline{J_\mathrm{ad}(\Omega)} = \frac{1}{2\pi}  |\sin \phi| \int _0^{2\pi} \d{\theta} \int _0^r r' \d{r'} F(\theta,r') \text{.}
\end{equation}

Let us first consider expanding around $\phi = \pi/2$; a straightforward integration for $N=2$ shows that
\begin{widetext}
\begin{align}
\overline{J_\mathrm{ad}(\Omega)} \simeq &\left\{-\frac{r^2w }{6[(2+w)^2-2r^2]^{3/2}} +  \frac{r^4w}{2[(2+w)^2-2r^2]^{5/2}}(\phi -\frac{\pi}{2}) 
-\frac{r^2w[11r^4+4r^2+(2+w)^2-(2+w)^4]}{12[(2+w)^2-2r^2]^{7/2}}(\phi -\frac{\pi}{2})^2 \right. \nonumber \\  & \; +\left.
\frac{r^4w[19r^4+16r^2+(2+w)^2-4(2+w)^4]}{12[(2+w)^2-2r^2]^{9/2}}(\phi -\frac{\pi}{2})^3
\right\}\Omega +O((\phi +\frac{\pi}{2})^4)\text{.}\label{adJ}
\end{align}
\end{widetext}
It can be confirmed that the adiabatic current is negative for all frequencies. Next, we expand the current for $N=2$ around $\phi = 0$, which gives
\begin{align}
\overline{J_\mathrm{ad}(\Omega)} \simeq &\left\{\frac{r^2w }{6[(2+w)^2-4r^2]^{3/2}} \phi \right. \nonumber \\ &+\left. \frac{r^2w[(2+w)^2+5r^2]}{36[(2+w)^2-4r^2]^{5/2}}\phi ^3 \right\}\Omega +O(\phi^5)\text{.}\label{geomJ0}
\end{align}
This indicates that in the absence of phase difference, no geometrical current is generated. While the above equations hold only for the adiabatic limit, the numerical results shown below support this conclusion also for the non-adiabatic regime.

\subsection{Numerical results}

To check the validity our analytical findings, we also numerically solve the Markovian ME, Eq. \eqref{best}, of the $N =2 $ system. The computations were performed for frequencies below $\Omega = 20$, and for a range of the system parameters $w$, $r$ and the phase difference $\phi$. It is found that to guarantee convergence and independence from initial values, $40$ initial cycles, i.e. $k_0 t_0 = 80\pi/\Omega$ is sufficient. Equations \eqref{numCurrdef} and \eqref{avgCurr} are used to compute the pumping current for each solution.

\begin{figure}
\centering
\includegraphics[width=0.5\textwidth]{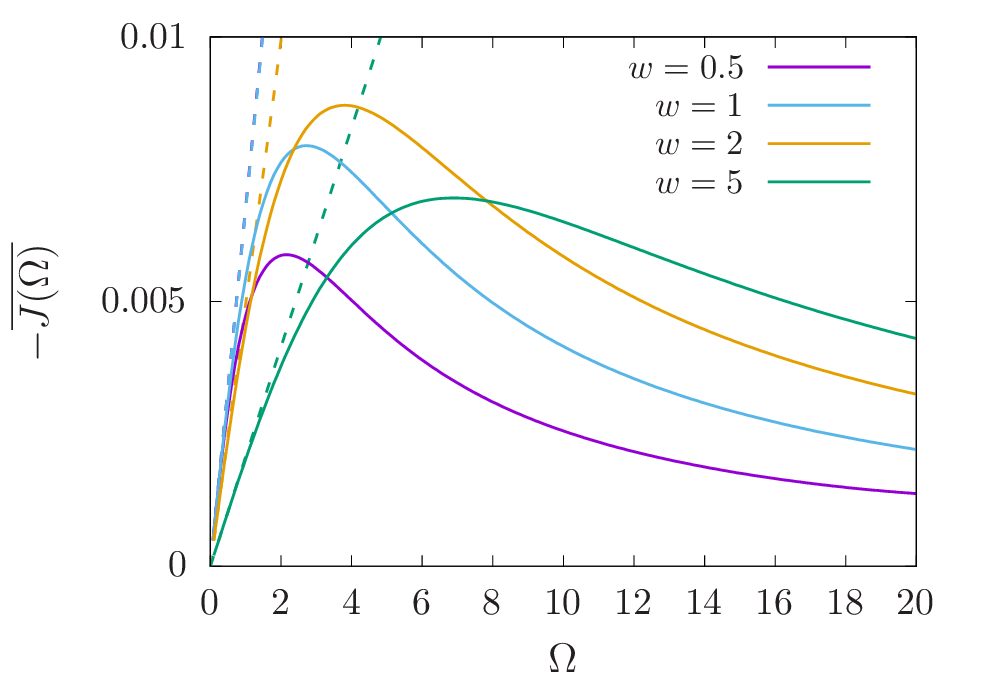}
\caption{\label{JR9} Averaged current $\overline{J(\Omega)}$ as a function of the modulation frequency for different values of $w$ at modulation amplitude $r=0.9$ and phase $\phi = \pi/2$,  obtained by solving Eq. \eqref{best} numerically. The dashed lines show the corresponding adiabatic approximations obtained from Eq. \eqref{adJ}.}
\end{figure}

We first show plots of the pumping current for selected values of $w$ with modulation amplitude $r=0.9$ and phase difference $\phi = \pi/2$ in Fig. \ref{JR9}. From these plots we see that the agreement between the numerical averaged current (solid lines) and the adiabatic current expressed by Eq. \eqref{adJ} (dashed lines) is good in the adiabatic regime $\Omega \to 0$.

\begin{figure}
\centering
\includegraphics[width=0.5\textwidth]{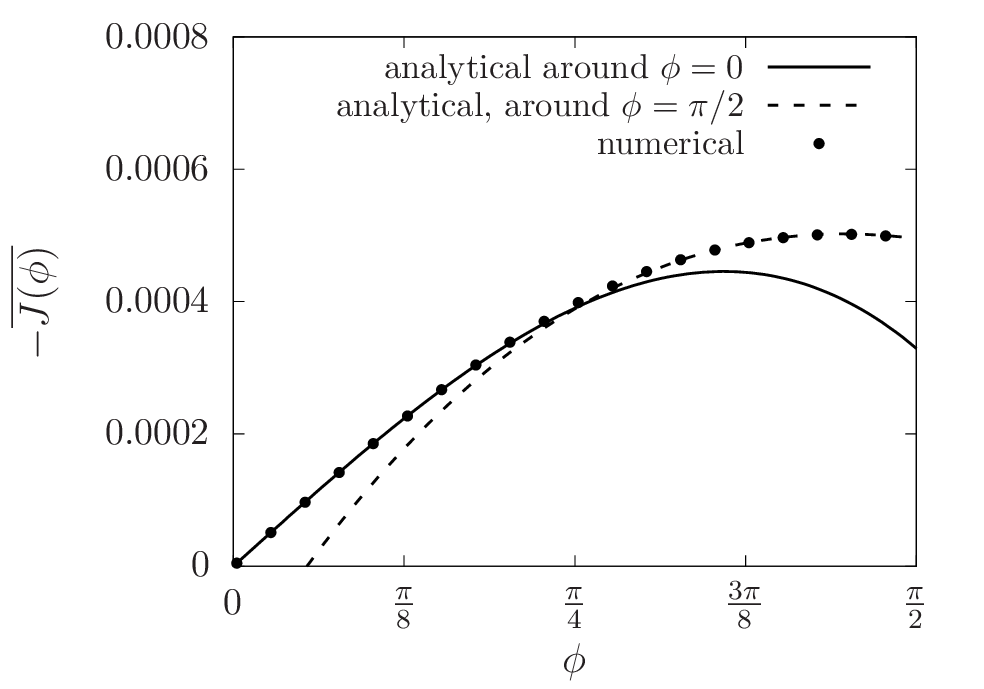}
\caption{\label{JP2} Averaged current $\overline{J(\phi)}$ as a function of the modulation phase difference $\phi$ with $r=0.9$ and $w=2$ and $\Omega = 0.1$. The solid circles represent the numerical current obtained by solving Eq. \eqref{best}, while the solid line corresponds to Eq. \eqref{geomJ0} (valid for small $\phi$), and the dashed line to \eqref{adJ} (valid for $\phi \simeq \pi /2$).}
\end{figure}

We have also investigated the dependence of the pumping current on the modulation phase difference $\phi$ in the low frequency regime ($\Omega =0.1$). Results are shown in Fig. \ref{JP2} , where the solid circles indicate the numerical result, obtained again by direct solution of Eq. \eqref{best}, the dashed lines correspond to the analytical result for $\phi \simeq \pi/2$ of Eq. \eqref{adJ} (including $O(\phi^3)$) and the solid lines show the result for $\phi \simeq 0$ as shown in Eq. \eqref{geomJ0} (including $O((\phi-\pi/2)^3)$). We see that both expansions agree well with the numerical results, and together the two expansions capture the $\phi$-dependence of the pumping current in the adiabatic regime sufficiently well.

\begin{figure}
\centering
\includegraphics[width=0.5\textwidth]{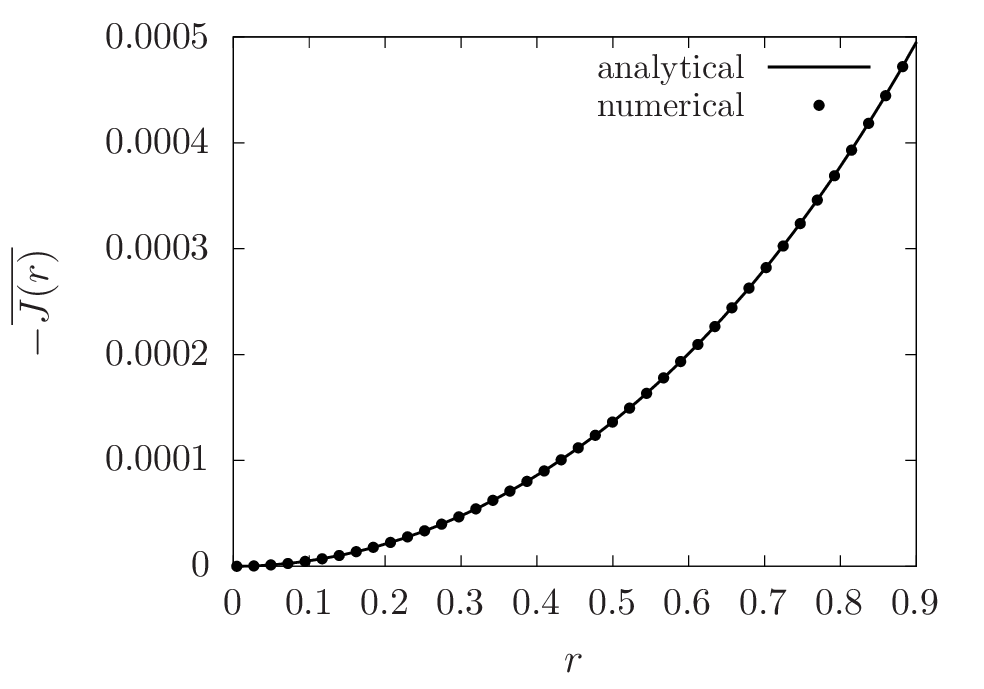}
\caption{\label{Jr} Averaged current $\overline{J(r)}$ as a function of the modulation amplitude  $r$ with $\Omega=0.1$, $\phi = \pi/2$ and $w=2$. The solid circles correspond to the numerical current obtained by solving Eq. \eqref{best} and the solid lines represent Eq. \eqref{adJ} .}
\end{figure}

\begin{figure}
\centering
\includegraphics[width=0.5\textwidth]{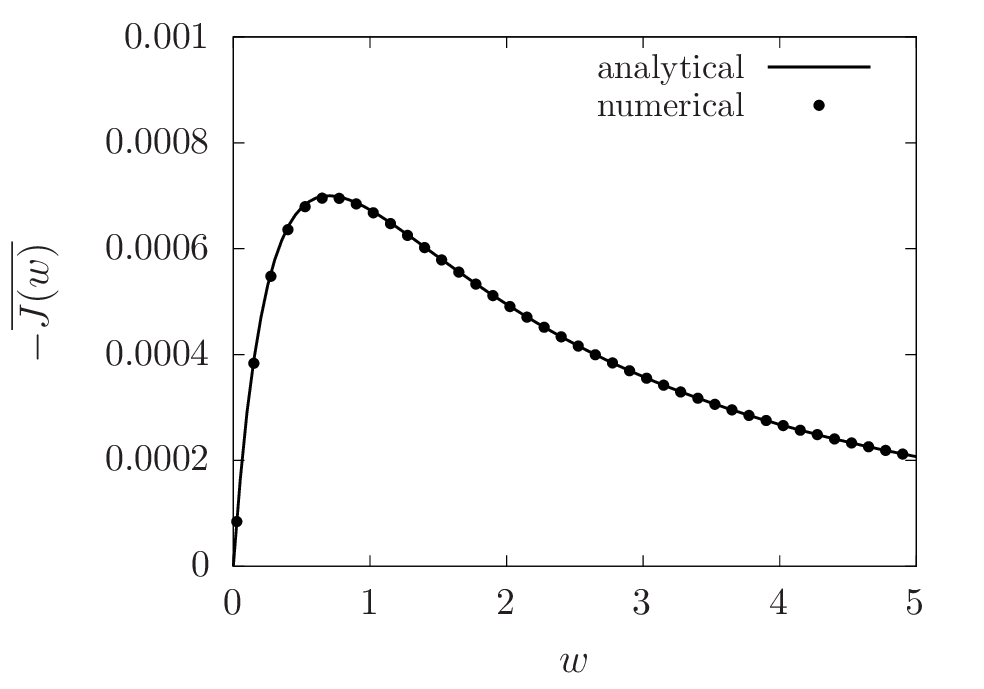}
\caption{\label{Jw} Averaged current $\overline{J(\phi)}$ as a function of the internal rate factor  $w$ with $\Omega=0.1$, $\phi = \pi/2$ and $r=0.9$. The solid circles correspond to the numerical current obtained by solving Eq. \eqref{best} and the solid lines represent Eq. \eqref{adJ}.}
\end{figure}

Finally, we investigate the dependence of the pumping current on the modulation amplitude $r$ and the internal rate factor $w$ again for the adiabatic regime ($\Omega =0.1$), as shown in Figs. \ref{Jr} and \ref{Jw}, respectively. It is again seen that in this regime, the agreement between the numerical and analytical results is good.

Comparing our results with those reported e.g. in Ref. \cite{fujii2019nonadiabatic}, we can see that the qualitative features of the current are unchanged. We note, however, that, due to the fact that our system is fundamentally different from the two-state Markovian SN model characterized only by a $2\times 2$ matrix, detailed comparison is not possible. 

We also note that the above analytical calculations can in principle be extended to the non-adiabatic regime by using a more general form of Eq. \eqref{adApp} that allows for transitions between different eigenstates. This approach is investigated in Appendix \ref{eigen}. The higher order eigenvectors have a complex form and the required integrations rapidly become involved. As shown in Appendix \ref{eigen}, we have confirmed the quantitative validity of our non-adiabatic analysis, at least at $O(\Omega^3)$. Compared to the ease of the above numerical treatment, however, these calculations do not yield any worthwhile insight into the physics, nor do they seem to offer a significant improvement over the adiabatic calculations.

\section{\label{discussion}Discussion}

The most important finding of our research is that a large class of systems possessing memory can be embedded into a larger Markovian system. The computational cost incurred due to the increased number of dynamical variables is outweighed by not having to deal with the time convolution integral, at least for the model we analyzed (see Eq. \eqref{subMatForm}). This is certainly true for numerical calculations, but we have shown that it also applies to the analytical calculations in the adiabatic limit.

Furthermore, analyzing the system in the non-Markovian framework does not have any apparent merits. Involving the time convolution integral in the discussion obscures the definition of the pumping current; indeed, it is not clear how to implement full counting statistics (FCS) in the presence of memory effects. On the other hand, since the larger Markovian system is not merely an approximation, but rather an exact equivalent of the non-Markovian system, reliability is not compromised by limiting oneself to the Markovian system. Thus, direct treatment of an equation of the type of Eq. \eqref{nMForm}  would be required only when the embedding into a larger Markovian system is not feasible.

\section{\label{conclusion}Conclusion}
In this paper, the Sinitsyn-Nemenman (SN) model, a periodically modulated $N$-state system, was generalized. It was shown that the non-Markovian SN model, governed by a non-Markovian master equation (nMME) which includes a time convolution integral, can be converted to a larger Markovian system, which is easier to analyze. Thus a method of solving the system dynamics at least numerically without needing to resort to any perturbative expansions was presented, yielding an approach to deal with this type of nMMEs which is in principle exact. In addition to solving the master equation numerically and using the results to calculate the pumping current for different values of the control parameters of the system, such as modulation frequency, phase and amplitude of the external modulation, we also presented an analytical calculation of the geometrical current in the adiabatic limit. It was found that the agreement between the numerical results and the analytical calculation was good.

Prospects of further research into this problem include the following: detailed microscopic derivation of the nMME in the framework of the present model, studying how the fluctuation theorem is affected by memory effects, and applying the method developed here to periodically driven quantum mechanical models. In particular, it would be interesting to see whether the dynamically modeled environment could be interpreted as the diagonal part of the density matrix of a two-state quantum system, thus leading to a connection between quantum coherence and non-Markovian time evolution. We are also interested in the extension of this analysis to a quantum mechanical system with charge current induced by the BSN curvature.
It should be noted that for such systems, the pumping charge current becomes zero if the repulsive interaction between electrons is not considered, but it has a finite value in the presence of such interactions \cite{PhysRevB.86.235308,yoshii2013analytical}. We also note that a quantum master equation can be derived by using Green's function technique \cite{PhysRevB.86.235308}.

\begin{acknowledgments}
The authors would like to thank Kazutaka Takahashi, Kazunari Hashimoto, Yuki Hino, Kiyoshi Kanazawa and Hiroyasu Tajima for fruitful discussions. This work is partially supported by Grant-in-Aid of MEXT for Scientific Research (Grant Nos. 16H04025 and 21H01006), a MEXT  scholarship and ISHIZUE 2020 of Kyoto University Research Development Program.
\end{acknowledgments}

\appendix

\section{\label{6x6Details}Details of the 6-state model}

In this Appendix, we consider the 6-state system (i.e. with $N=2$) obeying the Markovian ME of Eq. \eqref{best} in more detail, and outline the results used in the main text. 

\begin{figure*}
\includegraphics[width=\textwidth, trim={0 9cm 0 9cm}, clip]{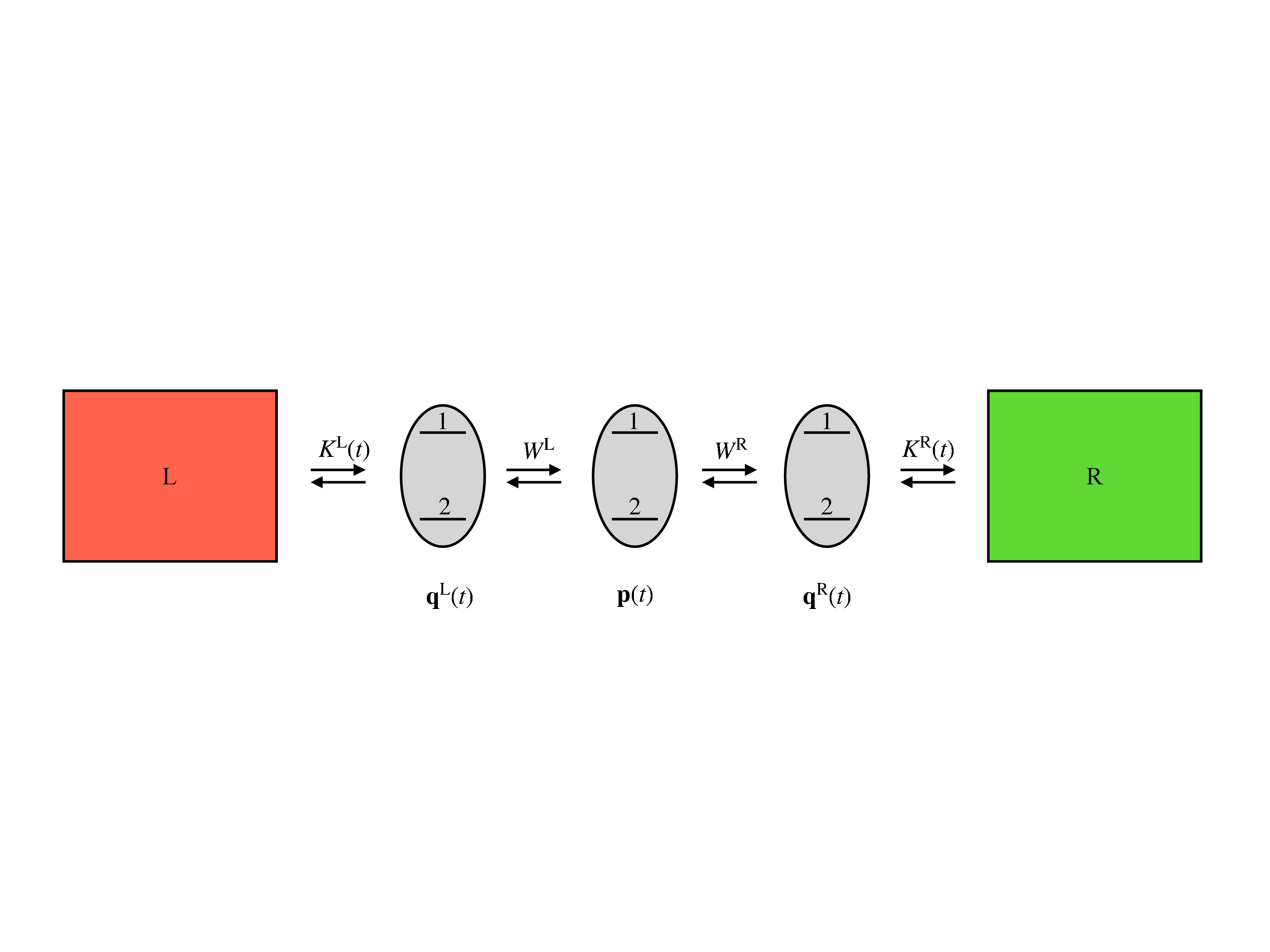}
\caption{\label{6x6one} A schematic of our 6-state Markovian model.}
\end{figure*}

We begin by obtaining the BSN curvature tensor using the machinery of full counting statistics (FCS)  \cite{Esposito_2009}. Firstly, we note that in vector notation, after adding the counting field $\chi$ to the transition rates between the right adjacent states and the right environment, the Markovian ME reads
\begin{equation}
\Omega \frac{\partial}{\partial{\theta}} \v{\rho}(\theta,\chi) = L(\theta,\chi) \v{\rho}(\theta,\chi) \text{,}
\end{equation}
where the full modified transition matrix $L(\theta,\chi) $ is written as
\begin{widetext}
\begin{equation}
L(\theta) = 
\begin{bmatrix}
-k_\mathrm{2} ^\mathrm{L} (\theta)- w &k^\mathrm{L}_\mathrm{1}(\theta) & w&  0&0&0 \\[0.5em] 
 k^\mathrm{L}_\mathrm{2}(\theta) &- k_\mathrm{1} ^\mathrm{L} (\theta)- w&0  &  w   &0&0 \\[0.5em] 
w&  0   &-2 w &0 &     w&  0  \\[0.5em] 
0&  w&0  &-2 w  &     0 &  w\\[0.5em] 
0&0& w&0&-k_\mathrm{2} ^\mathrm{R} (\theta)- w  &e^{\chi}k^\mathrm{R}_\mathrm{1}(\theta) \\[0.5em] 
0&0&0& w&e^{-\chi}k^\mathrm{R}_\mathrm{2}(\theta)&- k_\mathrm{1} ^\mathrm{R} (\theta)- w \\[0.5em] 
\end{bmatrix}\text{.}\label{fullglory}
\end{equation}
\end{widetext}
Figure \ref{6x6one} shows a schematic representation of this system: as described in the main text, there is no coupling between the two states at each site, except at the left and right reservoirs. Furthermore, we assume that only $ k_{1}^\nu$ and $ k_{2}^\nu$ depend on time.

We note here that, together with the observation that an alternative formula for the pumping current is 
\begin{equation}
J(\theta) = \sum _i  
\left. \frac{\partial }{\partial \chi} \right|_{\chi=0} L _{ij}(\theta,\chi)  \rho _i (\theta,0) \text{,}
\end{equation}
it can be seen from the above explicit form that unless $k^\mathrm{R}_\mathrm{2}(\theta) \neq k^\mathrm{R}_\mathrm{1}(\theta) $, the pumping current vanishes,  as mentioned in the main text below Eq. \eqref{param}. In more detail, having no phase difference between these hopping rates means that the cumulant generating function becomes $O(\chi ^2)$, resulting in the first derivative at $\chi =0$ evaluating to zero.

Let us now derive Eq. \eqref{BerryCurv}. In the adiabatic limit, we will only need to consider one eigenvalue, $\lambda(\theta,\chi)$ of the modified transition matrix,  and the corresponding right and left eigenvector $ \v{r} (\theta, \chi) $ and  $\boldsymbol\ell^\dag (\theta, \chi)$. This eigenvalue is taken to correspond to the zero-mode of the unmodified ME:
\begin{equation}
\lambda(\theta,0) = 0, \quad L (\theta, 0)  \v{r} (\theta, 0) = 0, \quad \boldsymbol\ell^\dag (\theta, 0)  L (\theta, 0) =0\text{.}
\end{equation}
In general, the above quantities  cannot be obtained exactly, but as we are only interested in their first derivatives with respect to $\chi$ here, we find approximate expressions
\begin{align}
L (\theta, \chi) &\simeq  L_0 (\theta) + L_1 (\theta) \chi\text{,} \\
\lambda(\theta,\chi) &\simeq \lambda _0(\theta) + \lambda _1 (\theta)\chi =  \lambda _1 (\theta)\chi\text{,}\\
\v{r} (\theta, \chi) &\simeq \v{r} _0(\theta) + \v{r} _1 (\theta)\chi\text{,} \\
\boldsymbol\ell^\dag (\theta, \chi) &\simeq \boldsymbol\ell^\dag _0(\theta) +\boldsymbol\ell^\dag _1 (\theta)\chi \text{.}
\end{align}
Since $\lambda (\theta, 0)=0$, we have the approximate eigenvalue $\lambda (\theta, \chi) \simeq   \lambda _1 (\theta)\chi$, which we wish to compute explicitly. We first note that the characteristic equation of the modified transition matrix for the eigenvalue $\lambda$ has the form 
\begin{align}
\det [L (\theta, \chi)  -\lambda \mathbb{1} ] &= a_n \lambda ^n +...+a_1 \lambda  + \chi  b_0 + O(\chi ^2) \nonumber \\ &= 0\text{.}
\end{align}
where $a_n$ are functions of the matrix elements of $L(\theta,0)$. This form guarantees that when $\chi =0$, $\lambda = 0$ is a solution of the characteristic equation. Thus, substituting in $\lambda  = \alpha \chi$ and ignoring all terms of $O(\chi ^2)$ or higher, we obtain $\lambda _1 = -b_0/a_1$. This eigenvalue approximation can then be used to obtain an approximation for the right and left eigenvectors. 
Writing down the eigenvalue equation to first order in $\chi$, we have
\begin{align}
[L_0 (t) + L_1 (\theta)&\chi][\v{r} _0(\theta) + \v{r} _1 (\theta)\chi]\text{,} \nonumber  \\ 
&\simeq  \chi L_0 (\theta)\v{r} _1 (\theta) + \chi L_1 (\theta) \v{r} _0(\theta)\text{,} \nonumber \\
&= \chi \lambda _1 (\theta) \v{r} _0(\theta)\text{,}
\end{align}
 and similarly for the left eigenvector. These systems of linear equations can be solved up to normalization algebraically. 
 
 Next, we need to impose the normalization $\boldsymbol\ell^\dag (\theta, \chi)  \v{r} (\theta, \chi) = 1$ up to first order in $\chi$, achieved by the normalization factor 
 \begin{equation}
 \v{r} (t, \chi) \mapsto N(\theta, \chi) \v{r} (\theta, \chi), \quad N(\theta, \chi) \simeq N_0(\theta) + N_1(\theta)\chi\text{.}
 \end{equation}
 We clearly must have
 \begin{align}
 N(\theta, \chi) &= [ \boldsymbol\ell^\dag (\theta, \chi)  \v{r} (\theta, \chi) ]^{-1} \nonumber \\
 &\simeq [\boldsymbol\ell^\dag _0(\theta)\v{r} _0(\theta) + \boldsymbol\ell^\dag _0(t)\v{r} _1 (\theta) \chi  +\boldsymbol\ell^\dag _1(\theta) \v{r} _0 (\theta)\chi ]^{-1} \nonumber \\
 &\simeq \frac{1}{\boldsymbol\ell^\dag _0(\theta)\v{r} _0(\theta)} -\frac{ \boldsymbol\ell^\dag _0(\theta)\v{r} _1 (\theta) +\boldsymbol\ell^\dag _1(\theta) \v{r} _0 (\theta)}{[\boldsymbol\ell^\dag _0(\theta)\v{r} _0(\theta)]^2} \chi \text{,}
 \end{align} 
 which shows that $N_0(\theta) = 1/[\boldsymbol\ell^\dag _0(\theta)\v{r} _0(\theta)]$ and $N_1(\theta) = -[\boldsymbol\ell^\dag _0(t)\v{r} _1 (\theta) +\boldsymbol\ell^\dag _1(\theta) \v{r} _0 (\theta)]/[\boldsymbol\ell^\dag _0(\theta)\v{r} _0(\theta)]^2$.
 
 Finally, we apply the above results to the calculation of the pumping current in the adiabatic limit. Assuming the normalization $\boldsymbol\ell^\dag (\theta, \chi)  \v{r} (\theta, \chi) = 1$, as long as $\Omega$ is sufficiently small, we can show that
\begin{equation}
\v{\rho}(\theta, \chi) \simeq e^{ \int _0 ^\theta\{ \frac{1}{\Omega} \lambda(\theta',\chi) - v(\theta',\chi)\} \d{\theta'}} \v{r} (\theta, \chi) \text{,}\label{adApp}
\end{equation}
where $v(\theta, \chi) := \boldsymbol\ell^\dag (\theta, \chi) \frac{\partial } {\partial \theta}\v{r}  (\theta, \chi) $ gives the geometrical contribution. Note that with our choice of modulation protocol, the dynamical term $ \lambda(\theta,\chi)$ averages to zero, and we are left with just $v(\theta,\chi)$. Using this approximation to compute the generating function of the current, and then picking out the term linear in $\chi$ (that is, taking the derivative with respect to $\chi$ and setting $\chi$ to zero), we obtain 
\begin{align}
\overline{J_\mathrm{ad}(\Omega)} &= - \frac{1}{2\pi}  \int _0^{2\pi}  \left. \frac{\partial }{\partial \chi} \right|_{\chi=0} v(\theta,\chi)  \d{\theta} \nonumber \\ 
&= \frac{\Omega}{2\pi} \oint _{\mathcal{C}(= \partial \mathcal{S})} A_\alpha  \d{k}^\alpha \label{geom1}\text{,}
\end{align}
where we performed a change of variables from time to the modulation parameter space spanned by $k_1$ and $k_2$. This is equivalent to Eq. \eqref{numCurrdef}. Here the Berry vector potential is defined as 
\begin{align}
A_\alpha (k_1, k_2) := \left. \frac{\partial }{\partial \chi} \right|_{\chi=0}   \boldsymbol\ell^\dag (k_1, k_2, \chi) \partial _{k_\alpha} \v{r}  (k_1, k_2, \chi)\label{geom2}\text{.}
\end{align}  
Finally, employing the generalized Stokes' theorem, which in the case of two-parameter modulation reduces to Green's theorem, we indeed obtain Eq. \eqref{BerryCurv}. This means that to obtain the adiabatic current, we need to compute $\lambda(\theta,\chi)$, $ \v{r} (\theta, \chi) $ and $\boldsymbol\ell^\dag (\theta, \chi)$ only, and only up to first order in $\chi$.

\section{\label{eigen}Eigenvalue expansion beyond the adiabatic limit}

In this section we consider an approach based on the full eigenvector expansion of the $N=2$ system \cite{fujii2019nonadiabatic}. We begin with obtaining all 6 eigenvalues and the corresponding right and left eigenvectors. Note that in this Appendix, the subscripts refer to the index of the eigenvalue or eigenvector, and series expansions are indexed by bracketed superscripts. Plots of the eigenvalues and each component of the right eigenvectors as a function of $\theta$ are shown for $w=2$, $\phi =\pi/2$, and $r=0.9$ in the figures below. We note that while for this choice of parameters, all the eigenvectors remain well-behaved for all $\theta$, exceptional points where the dimension of the vectors space spanned by the eigenvectors drops can arise for some modulation parameters. This may limit the usability of the approach presented here.
\begin{figure}
\centering
\includegraphics[width=0.5\textwidth]{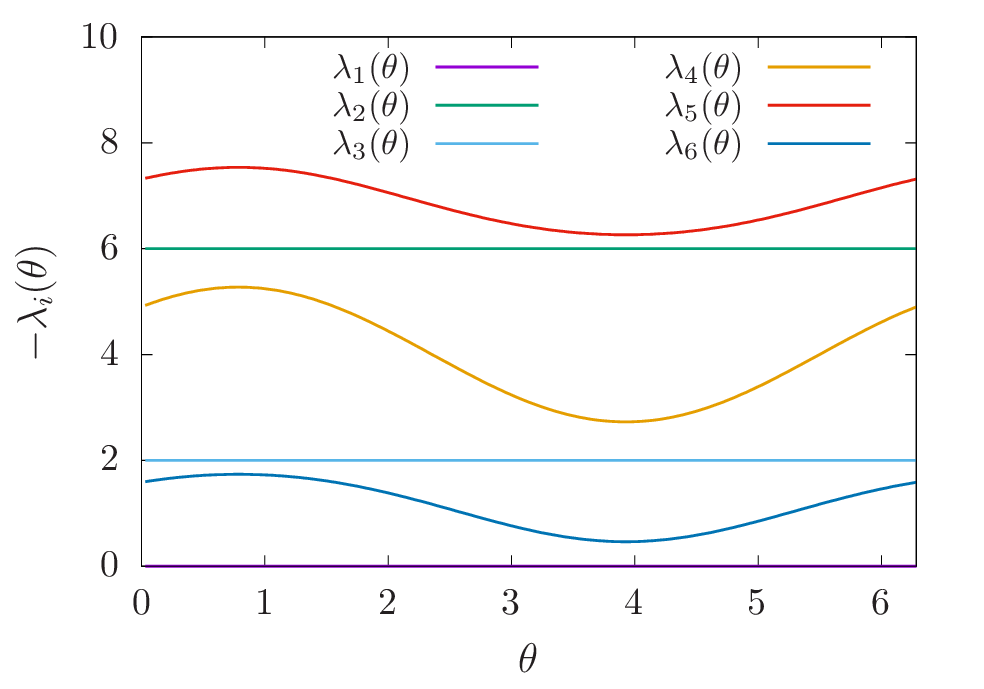}
\caption{\label{Lambda} Eigenvalues of the $N=2$ system plotted against $\theta$, with $r=0.9$, $\phi =\pi/2$ and $w=2$.}
\end{figure}

\begin{figure}
\centering
\includegraphics[width=0.5\textwidth]{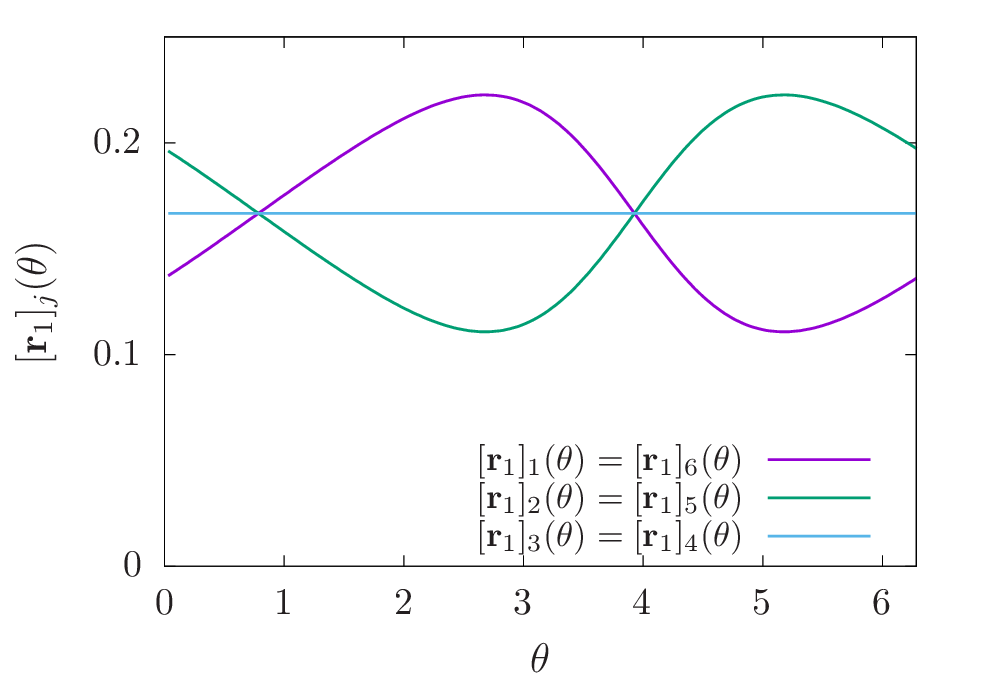}
\caption{\label{ER1} The $6$ components of the eigenvector $\v{r}_1$ of the $N=2$ system plotted against $\theta$, with $r=0.9$, $\phi =\pi/2$ and $w=2$.}
\end{figure}

\begin{figure}
\centering
\includegraphics[width=0.5\textwidth]{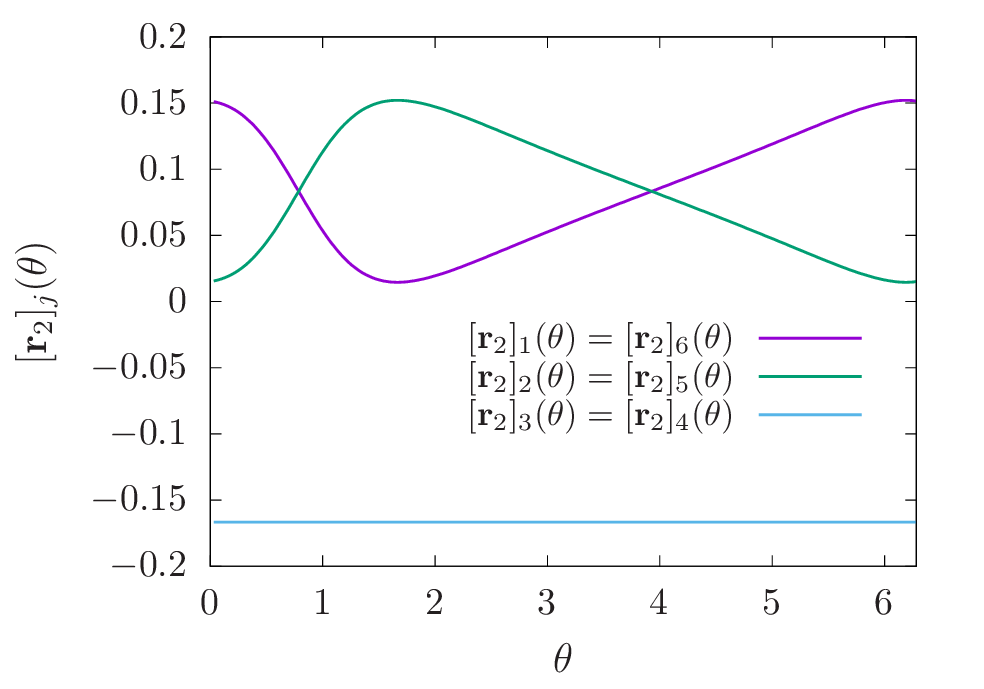}
\caption{\label{ER2} The $6$ components of the eigenvector $\v{r}_2$ of the $N=2$ system plotted against $\theta$, with $r=0.9$, $\phi =\pi/2$ and $w=2$.}
\end{figure}

\begin{figure}
\centering
\includegraphics[width=0.5\textwidth]{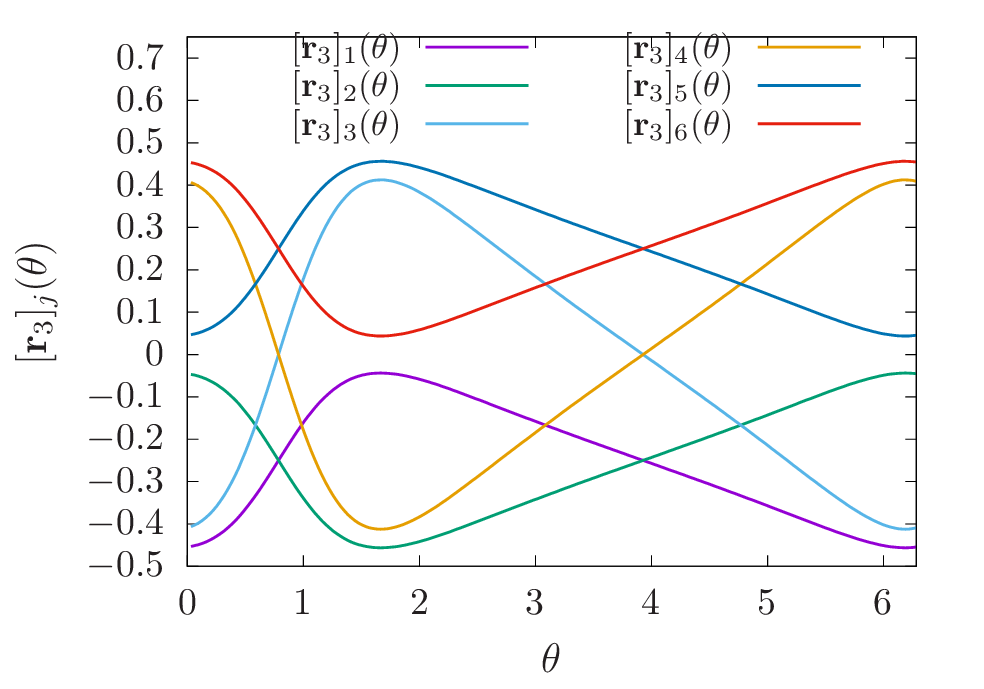}
\caption{\label{ER3} The $6$ components of the eigenvector $\v{r}_3$ of the $N=2$ system plotted against $\theta$, with $r=0.9$, $\phi =\pi/2$ and $w=2$.}
\end{figure}

\begin{figure}
\centering
\includegraphics[width=0.5\textwidth]{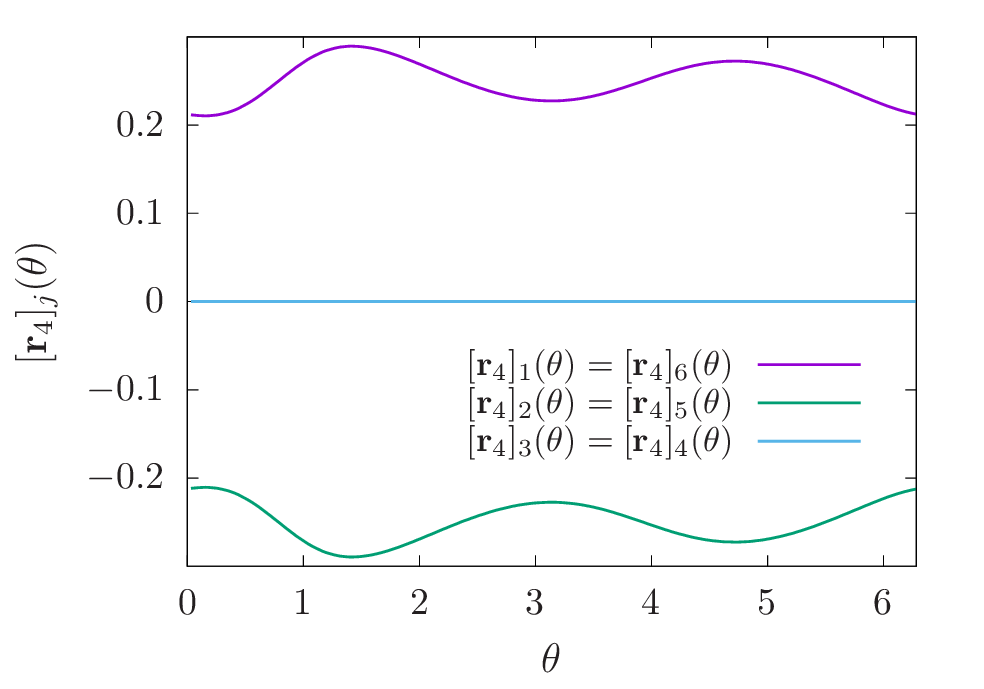}
\caption{\label{ER4} The $6$ components of the eigenvector $\v{r}_4$ of the $N=2$ system plotted against $\theta$, with $r=0.9$, $\phi =\pi/2$ and $w=2$.}
\end{figure}

\begin{figure}
\centering
\includegraphics[width=0.5\textwidth]{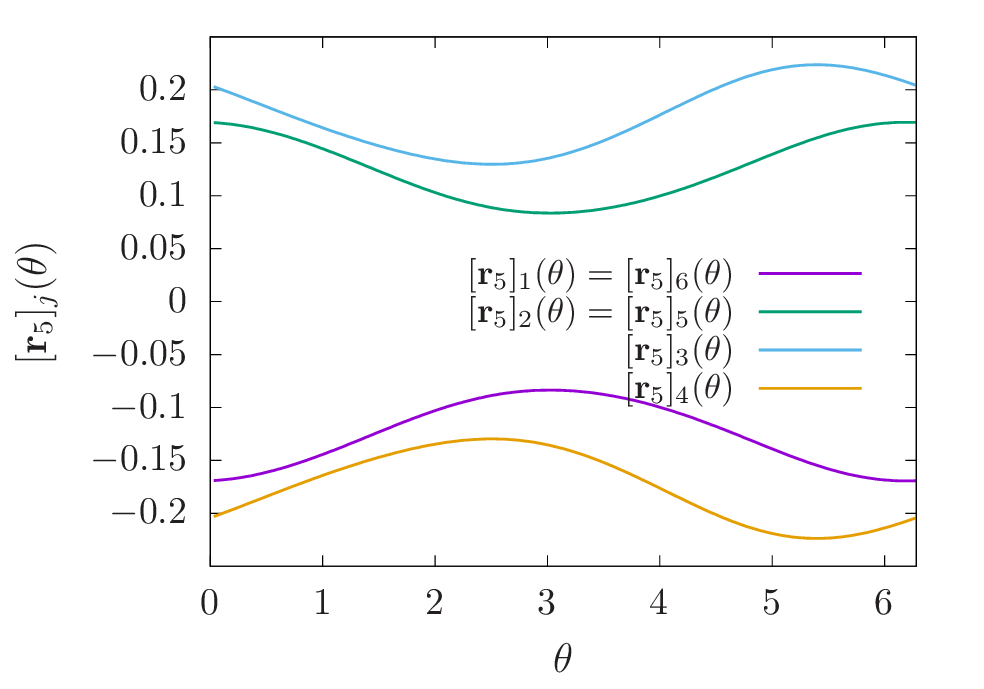}
\caption{\label{ER5} The $6$ components of the eigenvector $\v{r}_5$ of the $N=2$ system plotted against $\theta$, with $r=0.9$, $\phi =\pi/2$ and $w=2$.}
\end{figure}

\begin{figure}
\centering
\includegraphics[width=0.5\textwidth]{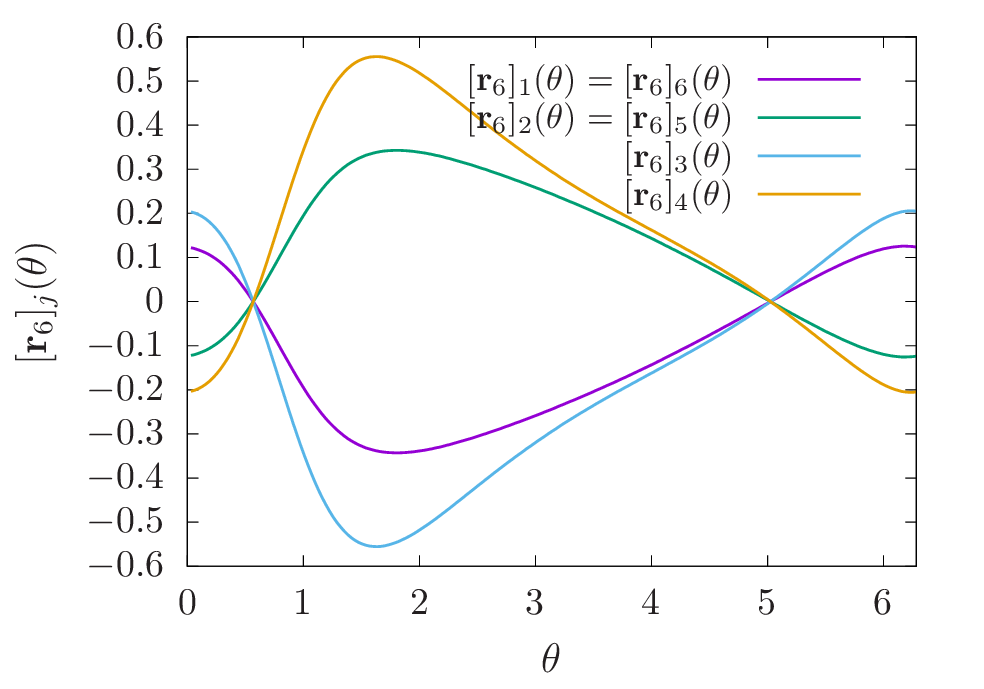}
\caption{\label{ER6} The $6$ components of the eigenvector $\v{r}_6$ of the $N=2$ system plotted against $\theta$, with $r=0.9$, $\phi =\pi/2$ and $w=2$.}
\end{figure}

\begin{figure}
\centering
\includegraphics[width=0.5\textwidth]{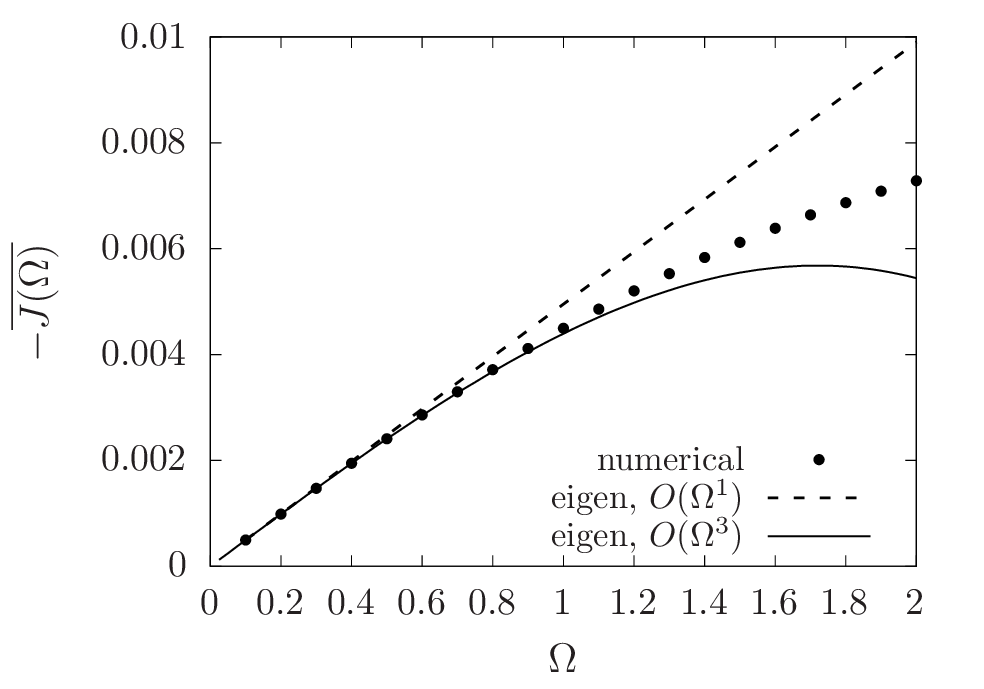}
\caption{\label{JPO} Averaged current $\overline{J(\Omega)}$ as a function of the modulation frequency  $\Omega$ with $\phi = \pi/2$, $r=0.9$ and $w=2$. The solid circles correspond to the numerical result obtained by solving Eq. \eqref{best}, while the dashed and solid lines represent the  $O(\Omega ^1)$ (adiabatic) and $O(\Omega ^3)$ results obtained by the eigenstate decomposition method of Appendix \ref{eigen}.}
\end{figure}
We normalize all eigenvectors as follows,
\begin{equation}
\boldsymbol\ell_i^\dag (\theta) \v{r}_j (\theta) = \delta_{ij}\text{,}
\end{equation}
and hence, necessarily we have
\begin{equation}
\sum_i \v{r}_i (\theta) \boldsymbol\ell_i^\dag (\theta) = \mathbb{1} \text{.}
\end{equation}
Transforming the eigenvectors according to
\begin{equation}
  \begin{cases}
    \tilde{ \v{r}}_i :=  e^{- \int_0^\theta  \boldsymbol\ell_i^\dag(\theta ')\dot{\v{r}}_i(\theta ')\d{\theta '}} \v{r} _i \text{,} \\[0.5em]
    \tilde{\boldsymbol\ell}_i :=  e^{ \int_0^\theta  \boldsymbol\ell_i^\dag(\theta ')\dot{\v{r}}_i(\theta ')\d{\theta '}}\boldsymbol\ell _i  \text{,} 
  \end{cases} \label{transform}
\end{equation}
where $\dot{\v{r}}:= \frac{\partial } {\partial \theta}\v{r}$, leaves orthonormality and completeness unchanged, but results in the convenient property
\begin{align}
\tilde{\boldsymbol\ell}_i^\dag \dot{\tilde{ \v{r}}}_i 
&= \tilde{\boldsymbol\ell}_i^\dag \left( e^{- \int_0^\theta  \boldsymbol\ell_i^\dag(\theta ')\dot{\v{r}}_i(\theta ')\d{\theta '}} \dot{\v{r}} _i 
- \boldsymbol\ell_i^\dag \dot{\v{r}}_i  e^{- \int_0^\theta  \boldsymbol\ell_i^\dag(\theta ')\dot{\v{r}}_i(\theta ')\d{\theta '}} \v{r} _i \right) \nonumber \\
&= \boldsymbol\ell_i^\dag \left(  \dot{\v{r}} _i 
- \boldsymbol\ell_i^\dag \dot{\v{r}}_i   \v{r} _i \right) 
= 0
\end{align}
Let us now expand the state vector in terms of the transformed eigenvectors:
\begin{equation}
\v{\rho} = \sum_{i=1}^6 c_i(\theta) e^{\frac{1}{\Omega}\int_0^\theta \lambda_i (\theta ')\d{\theta '}}  \tilde{ \v{r}}_i \text{.}
\end{equation}
Noting that since
\begin{equation}
\lambda_1  =  0 \text{,} \quad \boldsymbol\ell_1^\dag
\v{\rho} =1 \text{,}
\end{equation}
we see that $c_1(\theta) = e^{ \int_0^\theta  \boldsymbol\ell_i^\dag(\theta ')\dot{\v{r}s}_i(\theta ')\d{\theta '}}$, so that the expansion can be written as
\begin{equation}
\v{\rho} = \v{r}_1 + \sum_{i=2}^6 c_i(\theta) e^{\frac{1}{\Omega}\int_0^\theta  \lambda_i (\theta ')\d{\theta '}}  \tilde{ \v{r}}_i  \text{.}
\end{equation}
Substituting this into the original matrix equation, and multiplying on the left by  $\tilde{\boldsymbol\ell}_i$, $i>1$, we obtain
\begin{equation}
\dot{c}_i(\theta) e^{\int _0^\theta \lambda_i (\theta ')\d{\theta '}} +  \tilde{\boldsymbol\ell}_i^\dag\dot{{\v{r}}}_1 +
 \sum_{j\neq(1,i)}^6 c_j(\theta) e^{\frac{1}{\Omega}\int _0^\theta \lambda_j(\theta ')\d{\theta '}}  \tilde{\boldsymbol\ell}_i^\dag \dot{\tilde{ \v{r}}}_j = 0 \text{.}\label{approx}
\end{equation}
Neglecting the last term, we have for the expansion coefficients
\begin{equation}
c_i(\theta) \simeq C_i - \int_0^\theta  e^{-\frac{1}{\Omega}\int_0^{\theta '} \lambda_i(\theta '') \d{\theta ''}}  \tilde{\boldsymbol\ell}_i^\dag (\theta ')\dot{{ \v{r}}}_1 (\theta ') \d{\theta '}\text{,}\label{sln}
\end{equation}
so the expansion becomes
\begin{equation}
\v{\rho} \simeq
\v{r}_1 + \sum_{i=2}^6 \left\{ C_i e^{\frac{1}{\Omega}\int_0^\theta  \lambda_i(\theta ') \d{\theta '}} + \delta_i (\theta)\right\}\tilde{ \v{r}}_i\text{,}
\end{equation}
where
\begin{equation}
\delta_i (\theta) := -\int_0^\theta  e^{\frac{1}{\Omega}\int_{\theta '}^\theta  \lambda_i(\theta '') \d{\theta ''}}  \tilde{\boldsymbol\ell}_i^\dag (\theta ')\dot{{ \v{r}}}_1 (\theta ')\d{\theta '}\label{deltas}\text{,}
\end{equation}
whence
\begin{equation}
\delta_i =   \Omega  \frac{ \tilde{\boldsymbol\ell}_i^\dag \dot{{ \v{r}}}_1}{\lambda_i} + \Omega \frac{\dot{\delta}_i}{\lambda_i}\text{.}\label{deltaDE}
\end{equation}
Noting that the terms multiplying the constants of integration $C_i$ decay exponentially and can thus be neglected for long times, the expansion becomes
\begin{equation}
\v{\rho} \simeq
\v{r}_1 + \sum_{i=2}^6   \delta _i \tilde{ \v{r}}_i\text{.}\label{final}
\end{equation}
The numerical integrations required by Eq. \eqref{deltas} are rather slow, so let us obtain $\delta _i$ recursively instead. Defining $\delta _i^{(0)} :=  \Omega \frac{ \tilde{\boldsymbol\ell}_i^\dag \dot{{ \v{r}}}_1}{\lambda_i}$, we obtain 
\begin{equation}
\delta _i = \delta _i^{(0)}  + \sum _{k=1}^n \left(\frac{\Omega}{\lambda_i}  \frac{\d{}}{\d{\theta}} \right)^n  \delta _i^{(0)}\text{.}\label{derexp}
\end{equation}

To investigate the usefulness of this approach, we show the pumping current obtained for $n=0$, i.e. the adiabatic limit (solid line) and $n=2$, i.e. up to $O(\Omega ^3)$ (note that only odd powers of the frequency contribute to the current) compared to the numerical result from the main text in Fig. \ref{JPO}. We see that the adiabatic result is correctly reproduced by the eigenvalue analysis, and the addition of the $O(\Omega ^3)$ offers an improvement over the linear approximation.


\bibliography{Paper1}

\end{document}